\documentclass[12pt]{iopart}
\def\be{\begin{equation}}
\def\ee{\end{equation}}
\def\bear{\begin{eqnarray}}
\def\eear{\end{eqnarray}}

\def\bGamma{\mbox {\boldmath ${\Gamma}$}}

\include{boldsymb}

\def\bGamma{\mbox {\boldmath ${\Gamma}$}}

\begin{document}

\title{Structure of multi-component/multi-Yukawa mixtures
}
\author{
L. Blum\dag 
\footnote[3]{To
whom correspondence should be addressed
 (lesblum@yahoo.com)}
  and M. Arias\dag  
}

\address{\dag Department of Physics, P.O. Box 23343, University of Puerto Rico, Rio Piedras, PR 00931-3343 
and Department of Mathematics, Hill Center Rutgers University, Piscataway N.J. 08854 }

\begin{abstract}
Recent small angle scattering experiments
reveal new  peaks  in the structure function $S(k)$ of colloidal systems ( S.H. Chen et al \cite{chen05}), in a region that was inaccessible with older instruments.
 It has been increasingly evident that a single (or double) Yukawa MSA-closures   cannot account for these observations, and three or more terms are needed. On the other hand the MSA is not sufficiently accurate\cite {broccio06} more accurate theories such as the HNC have been tried.
But while the MSA is asymptotically exact at high densities \cite{yrlb1}it does not satisfy the low density asymptotics. This has been corrected in the soft-MSA \cite {blnart72,blnart74}  by adding exponential type terms. The results compared to experiment and simulation  for liquid sodium by Rahman and Paskin ( as shown in \cite {blnart72}) are remarkably good. 
 We use here  a general closure  of the Ornstein Zernike equation  which is not necessarily  the MSA closure \cite{bhe01}.
\be
c_{ij}(r)
=\sum_{n=1}^{M}{\cal K}_{ij}^{(n)}e^{-z_{n}r}/r\label;\qquad{\cal K}_{ij}^{(n)}=K^{(n)}\delta
_{i}^{(n)}\delta _{j}^{(n)};\qquad r\geq \sigma_{ij}
\label{eq1}
\ee
with the boundary condition for $g_{ij}(r)=0$ for $r \leq  \sigma_{ij}$. 
This general closure  of the Ornstein Zernike equation will go well beyond the MSA since it has been tested by Monte Carlo simulation   for tetrahedral water  \cite{bluvede}, toroidal ion channels  \cite{blen05}  and  polyelectrolytes \cite{blube04}. For this closure we get  for the Laplace transform of the pair correlation function an {\it explicitly } symmetric result
\bear
2 \pi {\tilde g}_{ij}(s)&=&-\frac{e^{-s \sigma_{ij}}}{D_\tau(s)}\left\{{1\over s^2}+ {1\over s}Q'_{ij}(\sigma_{ij})+\sum_{m=1}^{M}{{ z_m\tilde {\mathcal  X}}_i^{(m)}{f}_j^{(m)}\over{s+z_m}}\right\}\nonumber \\
\label{eq2}
\eear
 This function is also easily transformed into $S(k)$ by replacing $s\Rightarrow ik$. For low density situations (dilute colloids) $D_\tau (s)\sim 1+{\mathcal O(\rho)}$ and the $S(k)$ is  a sum of  $M$ Lorentzians.
For hard sphere PY mixtures we get the simple (compare \cite{lebo64,bs79})
\[
2 \pi {\tilde g}_{ij}(s)=-\frac{e^{-s \sigma_{ij}}}{s ^2 D_\tau(s)}\left\{1+s\left[(Q^{HS})'_{ij}(\sigma_{ij})\right]\right\}
\]
where $D_\tau (s)$ is a scalar function.
For  polydisperse electrolytes in the MSA a simpler  expression  is also obtained (compare \cite{bh77}) .
An explicit continued fraction solution of the 1 component multiyukawa case is also given.\\

\end{abstract}
{\tiny\today , yuksowhsinv3a, 
ArXiv cond-mat/0602477, 
IOP : CM/219099/SPE/8462}\\
\submitto{\JPCM}
\maketitle

\section{\label{sec:level1}Introduction}
 There are many problems of practical and academic interest
that can be formulated as closures of some kind of either scalar or matrix Ornstein-Zernike (OZ) equation. These closures can always be expressed by a sum of exponentials, which do form a complete basis set if we allow for complex numbers \cite{blmhe3,blmub}. It is of practical interest to be able to relate small angle scattering experiments as directly as posible to theoretical (molecular) parameters.
For a number of systems the MSA (\cite {blmhe3,py,lebperc66})has been  generally adequate (\cite{chen05,lin02}). The GOCM which is a the single peak-single Yukawa closure-MSA description  is adequate in the simple cases.
However, recent high resolution  experiments have shown that to account for the additional peaks  seen in the SANS experiments more yukawas are required \cite{broccio06,ff9,ff10}.
The MSA and  HNC closures of the  OZ  equation  slightly underestimate the height of the interaction peak
in the structure factor, whereas a close comparison of the
cluster peak position is somewhat hindered by the limited resolution in Q of the simulation data, because  the largest  simulation
box is never big enough. In most part of the practical
range of potential parameters, the theoretical predictions
coming from the two closures  have
comparable accuracy. Therefore, for  fitting  the          
neutron scattering intensity distributions,  analytical structure factors  are
preferable since the resolution of the equations is more stable and fast.

Our present solution makes  the direct comparison between theory and experiment  feasible since a simple expression for the Fourier transform $S(Q)$ is proposed which is directly related to the closure of the OZ equation.  In fact with the very mild assumption made is that the direct correlation function can be expanded as

\be
c_{ij}(r)=\sum_{n=1}^{M}K_{ij}^{(n)}e^{-z_{n}(r-\sigma _{ij})}/r 
=\sum_{n=1}^{M}{\cal K}_{ij}^{(n)}e^{-z_{n}r}/r;\quad \quad r >  \sigma_{ij}\label {eq3} 
\ee
In the factorizable case (which is also the electrostatic charge case), we have 
\be
{ K}_{ij}^{(n)}=K^{(n)}\delta
_{i}^{(n)}\delta _{j}^{(n)};
\quad {\cal K}_{ij}^{(n)}=K^{(n)}d_{i}^{(n)}d_{j}^{(n)};\quad \delta _{i}^{(n)}=d_{i}^{(n)}e^{-z_{n}\sigma _{i}/2}\label{eq4} 
\ee
with \[g_{ij}(r)=0\quad for \quad r \leq  \sigma_{ij}.\]

 Here $z_n$ can be a complex number, and then eq.(\ref{eq3}) is a complete function basis set. This closure works very well with a number of potentials, that include Lennard-Jones, Buckingham, and liquid metals \cite{blnart72,bh78}. The effective interaction among charged colloids and globular proteins in solution is also well described by this
potential which is the sum of a short range part attraction, usually accounting for van der Waals or entropic forces, plus
a long range repulsion (of up to a few particle diameters),
arising from screened electrostatic potential\cite{ffLR}.
The two-Yukawa (2Y) fluid, consisting of particles interacting with this potential can be usefully applied to a wide variety of systems such as C60 fullerenes at high
temperature\cite{ff4}, globular proteins\cite{ff2,ff3}, and the Derjaguin-Landau-
Verwey-Overbeek (DLVO)\cite{ff1} potential for liophobic colloids.

While the initial motivation was to study simple approximations like the
Mean Spherical (MSA) or Generalized Mean Spherical Approximation (GMSA), the
availability of closed form solutions for the general closure of the hard
core OZ equation makes it possible to write down analytical solutions for
any given approximation that can be formulated by writing the direct
correlation function $c(r)$ outside the hard core as

\be
c(r)=\sum_{n=1}^{M}K^{(n)}e^{-z_{n}(r-\sigma )}/r 
=\sum_{n=1}^{M}{\cal K}^{(n)}e^{-z_{n}r}/r\label
{eq:yk1} 
\ee
 In this equation $K^{(n)}$ is the interaction/closure constant 
used in the general solution  first found by Blum and Hoye (which we 
will call BH78) \cite
{bh78}, while ${\cal K}^{(n)}$ is the definition used in the later 
general solution by Blum, Vericat and Herrera (BVH92 in what follows) 
\cite{bvh92}. In this work we will use the more common notation of BVH92. 
The case of factored interactions was discussed by Blum, \cite{b80,b3} and by Ginoza \cite{gin1,gin2,gin5,gin6}. 
The general solution of this problem  \cite{bvh92} is given in terms of the scaling matrix $
\bGamma $ which will comply with the physical constraints of the systems (Blum et al. \cite
{blmub,blmhe1,bluhe02}). The solution of the resulting algebraic equations has a number of branches (Pastore \cite{pastore88}). The rigorous analysis of this question is very complex. We have used the following working hypothesis:
\begin{enumerate}
\item The singularities of the equations are poles, branch cuts and essential ( exponential) sigularities.
\item The physical branch  is the one for which the correct zero density result is obtained. This is equivalent to what is done in the rigorous treatment van der Waals theory, and  includes metastable regions and spinodals.
\item The analysis of the repulsive part of the potential is more delicate since it is clearly related to the exponential singularities, and the convexity  of the limiting high density configurations \cite{hernando93}.
\end{enumerate}

 For only one component the matrix $
\bGamma $ can be assumed  to be diagonal without loss of generality, and explicit expressions for the closure relations
for any arbitrary number of Yukawa exponents $M$ were obtained. The
solution is  then remarkably simple in the MSA since then explicit formulas for
the thermodynamic properties are obtained. However there are undetermined integration constants in the entropy which have to be adjusted to get the correct limiting behavior \cite{lin02}.

\section{Summary of Previous Work}
\label{prework}

We study the Ornstein-Zernike (OZ) equation 
\be
h_{ij}(12)=c_{ij}(12)+\sum_{k}\int d3h_{ik}(13)\rho _{k}c_{kj}(32)\label
{eq:oz} 
\ee
where $h_{ij}(12)$ is the molecular total correlation function and $%
c_{ij}(12)$ is the molecular direct correlation function, $\rho _{i}$ is the
number density of the molecules i, and $i=1,2$ is the position $\vec{r}_{i}$
, $r_{12}=|\vec{r}_{1}-\vec{r}_{2}|$ and $\sigma _{ij}$ is the distance of
closest approach of two particles $i,j$. The direct correlation function is

\be
c_{ij}(r)=\sum_{n=1}^{M}K_{ij}^{(n)}e^{-z_{n}(r-\sigma _{ij})}/r,\qquad
r>\sigma _{ij}\label{eq:msac} 
\ee

\noindent and the pair correlation function is 
\be
h_{ij}(r)=g_{ij}(r)-1=-1,\qquad r\leq \sigma _{ij}\label{eq:msac1} 
\ee

We use the Baxter-Wertheim (BW) factorization of the OZ
equation

\be
\left[ {\bf I}+{\bf {\rho }{H}(k)}\right] \left[ {\bf I}-{\bf \rho }%
{{\bf C}}(k)\right] ={\bf I}\label{eq:oz2} 
\ee
where $I$ is the identity matrix, and we have used the notation 
\be
{{\bf H}}(k)=2\int_{0}^{\infty }dr\cos (kr){\bf J}(r)\label{eq:oz2a} 
\ee
\be
{{\bf C}}(k)=2\int_{0}^{\infty }dr\cos (kr){\bf S}(r)\label{eq:oz2b} 
\ee

The matrices $J$ and $S$ have matrix elements

\be
J_{ij}(r)=2\pi \int_{r}^{\infty }dssh_{ij}(s)\label{eq:oz3a} 
\ee
\be
S_{ij}(r)=2\pi \int_{r}^{\infty }dssc_{ij}(s)\label{eq:oz3b} 
\ee

\be
\left[ {\bf I}-{\bf \rho }{{\bf C}}(k)\right] =\left[ {\bf I}-{\bf 
\rho }{{\bf Q}}(k)\right] \left[ {\bf I}-{\bf \rho }{{\bf Q}}
^{T}(-k)\right] \label{eq:factor} 
\ee
where ${{\bf Q}}^{T}(-k)$ is the complex conjugate and transpose of $
{{\bf Q}}(k)$. The first matrix is non--singular in the upper half
complex $k$-plane, while the second is non--singular in the lower half
complex $k$-plane.

It can be shown that the factored correlation functions must be of the form 
\be
{{\bf Q}}(k)={\bf I}-{\bf \rho }\int_{\lambda _{ji}}^{\infty }dre^{ikr}
{{\bf Q}}(r)\label{eq:q(k)q(r)} 
\ee

\noindent where we used the following definition 
\be
\lambda _{ji}={\frac{1}{2}}(\sigma_j -\sigma _{i})\label{eq:q7} 
\ee

{ 
\be
{\bf S}(r)={\bf Q}(r)-\int dr_{1}{\bf Q}(r_{1})\rho {\bf Q^{T}}(r_{1}-r)\label{eq:sq(r)} 
\ee
}

Similarly, from Eq. (\ref{eq:factor}) and Eq. (\ref{eq:oz2}) we get, using
the analytical properties of $Q$ and Cauchy's theorem 
\be
{\bf J}(r)={\bf Q}(r)+\int dr_{1}{\bf J}(r-r_{1}){\bf \rho }{\bf Q}(r_{1})%
\label{eqjq(r)} 
\ee

The general solution is discussed in \cite{b80,gin1,bluhe02}, and yields

\be
q_{ij}(r)=q_{ij}^{0}(r)+\sum_{n=1}^{M}D_{ij}^{(n)}e^{-z_{n}r}\qquad \lambda
_{ji}<r\label{eq:p1} 
\ee

\[
q^0_{ij}(r)=(1/2)A_{j} [(r-\sigma_j/2)^2-(\sigma_i/2)^2] +
\beta_j[(r-\sigma_j/2)-(\sigma_i/2)]
\]
\be
+\sum_{n=1}^{M}C^{(n)}_{ij}e^{-z_n \sigma_{j}/2}
[ e^{-z_n (r-\sigma_j/2)} -e^{-z_n \sigma_{i}/2}]
\qquad \lambda_{ji}<r<\sigma_{ij}\\
\label{eq22d}
\ee

The parameters $A_j,\beta_j, C_{ij}, D_{ij} $ are defined below and in  Appendix 1.\\

\section{ The Laplace Transforms and the structure functions}

From Eq. (\ref{eqjq(r)}) we obtain
the Laplace transform of the pair correlation function 
\be 
2 \pi\sum_{\ell} {\tilde g}_{i{\ell}}(s) [\delta_{\ell j}- \rho_{\ell} {\tilde q}_{{\ell}j}(i s)]=
{\tilde q}^{0^{'}}_{ij}(i s)
\label{eqp22}
\ee
where  
\[ 
\tilde{q}^{0^{'}}_{ij}(i s)=\int_{\sigma_{ij}}^{\infty} dr 
e^{-s r}[q^0_{ij}(r)]'
\]
\be
\qquad =[(1+s \sigma_i/2) A_j+s \beta_j]e^{-s \sigma_{ij}}/s^2-
\sum_m \frac{z_m}{s+z_m} e^{-(s+z_m) \sigma_{ij}}C^{(m)}_{ij}
\label{eqp23}
\ee
and
\bear 
& & e^{s \lambda_{ji}}\tilde{q}_{ij}(i s)=
\sigma_i^3\psi_1( s\sigma_i)A_j +\sigma_i^2\phi_1( s\sigma_i)\beta_j + \\ \nonumber
& &\sum_m \frac{1}{s+z_m}[(C_{ij}^{(m)}+ D_{ij}^{(m)})e^{-z_m \lambda_{ji}}
-C_{ij}^{(m)}e^{-z_m \sigma_{ji}}  - z_m \sigma_i\phi_0( s\sigma_i)C_{ij}^{(m)} e^{-z_m \sigma_{ji}}]
\label{eqp24}
\eear
where\\
\bear
\psi _{1}(x)&=&[1-x/2-(1+x/2)e^{-x}]/(x^{3})=\frac{\sigma^2}{2 z} e^{-z \sigma/2} i_1(z \sigma/2)
 \nonumber \\
i_1(x)&=&{1\over x^2}(\sinh {x}-x \cosh{x}) \nonumber \\
\phi _{1}(x)&=&[1-x-e^{-x}]/(x^{2})=x\psi _{1}(x)-\phi _{0}(x)/2;\qquad 
\phi _{0}(x)={[1-e^{-x}]\over x}\nonumber \\
\label{eq26}
\eear
In the factored case (see eq(\ref{eq4}))
\bear
D_{ij}^{(m)}=-\delta_i^{(m)} a_j^{(m)};\quad C_{ij}^{(m)}=\left(\delta_i^{(m)}-\frac{{\mathcal B}_i^{(m)}}{z_m}\right)a_j^{(m)};\quad
{\mathcal B}_i^{(m)}=2 \pi \sum_k \rho_k g_{ik}^{(m)} \delta_k^{(m)}
\nonumber \\
\eear
We remember that \cite{bvh92}

\bear
 2 \pi \sigma_{ij}g_{ij}(\sigma_{ij})&=&q_{ij}^{'}(\sigma_{ji}^{-})-q_{ij}^{'}(\sigma_{ji}^{+})\nonumber \\&
=&A_{j} (\sigma_i/2) +
\beta_j
-\sum_{m=1}^{M} \left(z_m\delta_i^{(m)}-{\mathcal B}_i^{(m)}\right)a_j^{(m)} e^{-z_m \sigma_{ij}}
\label{eqq26}
\nonumber \\
\eear
and
\bear 
\tilde{q}_{ij}(i s)&=&
e^{-s \lambda_{ji}}\sigma_i^3\psi_1( s\sigma_i)A_j +e^{-s \lambda_{ji}}\sigma_i^2\phi_1( s\sigma_i)\beta_j -  \nonumber \\
 \sum_{m} \frac{e^{-z_m \lambda_{ji}}}{s+z_m}&  &\left(\delta_i^{(m)}+(\hat{\cal B}_i^{(m)}-\delta_i^{(m)})\left\{1-e^{-(z_m +s)\sigma_{i}}\left[1
+z_m \sigma_i\phi_0(\sigma_i s)\right]\right\}\right)a_j^{(m)}
\nonumber \\
\label{eqp26a}
\eear
which can be written in the form
\bear
 {{\mathbf M }_{ij}{(s)}}=\delta_{ij}&-&\rho_i {q}_{ij}(i s)\nonumber \\
& = &\delta_{ij}-a_i b_j-c_i d_j -\sum_m^M e^{(m)}_i f^{(m)}_j \equiv\sum_{\alpha=1}^{M+2}{ a_i^{(\alpha)}}{ b_j^{(\alpha)}}
\label{eqp30}
\eear
with
\bear
a_i&=&\rho_i \sigma_i^3 \psi_1(s \sigma_i)e^{s\sigma_i/2}\qquad b_j=A_je^{-s\sigma_j/2}=A_{j}^{0}e^{-s\sigma_j/2}+\frac{\pi}{\Delta}\sum_{n}a_{j}^{(n)}P^{(n)}e^{-s\sigma_j/2}\nonumber \\
c_i&=&\rho_i \sigma_i^2 \phi_1(s \sigma_i)e^{s\sigma_i/2}\qquad d_j=\beta_je^{-s\sigma_j/2} =\beta_{j}^{0}e^{-s\sigma_j/2}+\frac{2\pi}{\Delta}\sum_{n}a_{j}^{(n)}\Delta
^{(n)}e^{-s\sigma_j/2}\nonumber \\
e^{(m)}_i&=&\rho_i e^{s\sigma_i/2}\sigma_i^2 \varphi^{(m)}_1(s \sigma_i)\qquad f^{(m)}_j=a^{(m)}_j e^{z_m \sigma_j/2}\nonumber \\
 \varphi^{(m)}_1(s \sigma_i)&=& \frac{e^{z_m \sigma_i/2}}{s+z_m}\left\{\frac{{\mathcal B}_i^{(m)}}{z_m}-e^{-z_m \sigma_{i}}\left(\frac{{\mathcal B}_i^{(m)}}{z_m}-\delta_i^{(m)}\right)\left[1
+\frac{z_m}{s}(1-e^{-s\sigma_i})\right]\right\}\nonumber \\
\label{eqp31}
\eear
Notice that eq.(\ref{eqp23}) can be written as
\bear
\tilde{q}^{0^{'}}_{ik}(i s)=-{\widetilde  a_i}{ b_k}-{\widetilde c_i}{ d_k}-\sum_m^M {\widetilde e_i}^{(m)}{ f_k^{(m)}} \equiv {\widetilde \mu}_{ij}=\sum_{\alpha=1}^{M+2}{\tilde a_i^{(\alpha)}}{ b_j^{(\alpha)}}\nonumber \\
\eear
with
\bear
\tilde{\bf a}_i&=&{(1+s \sigma_i/2)\over s^2} e^{-{s \sigma_{i}\over2}}\nonumber \\
\tilde{\bf c}_i&=&{e^{-{s \sigma_{i}\over2}}\over s}\nonumber \\
\tilde{\bf e}^{(m)}_i&=&e^{-{s \sigma_{i}\over2}}\frac{z_m}{s+z_m}{\mathcal X}_i^{(m)};\quad {\mathcal X}_i^{(m)}\equiv e^{-z_m \sigma_{i}}\left(\frac{{\mathcal B}_i^{(m)}}{z_m}-\delta_i^{(m)}\right)\nonumber \\
\label{eqp33}
\eear
With this notation we can rewrite eq.(\ref{eqp22}) as
\be
2 \pi {\tilde g}_{ij}(s) =\sum_k {\widetilde \mu}_{ik}\left[{\mathcal M }_{ij}(s)\right]^{-1}
\ee
After some algebra we get the general  and simple result
\bear
2 \pi {\tilde g}_{ij}(s)&=&-\frac{\widetilde \mu_{ij}(s)}{D_\tau(s)}
\nonumber \\
\label{eq35n}
\eear
with\\
\bear
D_\tau=Det\left|\delta_{\alpha,\beta}-\left(\sum_i a_i^{(\alpha)} b_i^{(\beta)}\right)\right| ;\quad 
\eear
\section{Selected applications}

We now apply the  general result eq.(\ref{eq35n}) to a few selected examples of interest:

\subsection{The Polydisperse hard spheres mixture}
For hard spheres
\bear
A_j&=&\frac{2\pi}{\Delta}\left[1+\frac{ \pi \zeta_2}{2\Delta}\sigma_j\right]=\frac{2\pi}{\Delta}\left[1+\frac{  \zeta_2}{2}\beta_j\right]
\nonumber \\
\beta_j&=&\frac{ \pi }{\Delta}\sigma_j
\nonumber 
\eear
Then we get, using the definitions of eq.(\ref{eqp30}-\ref{eqp33})
\bear
a_i&=&\rho_i \sigma_i^3 \psi_1(s \sigma_i)\qquad b_j=A_j^0\nonumber \\
c_i&=&\rho_i \sigma_i^2 \phi_1(s \sigma_i)\qquad d_j=\beta_j^0 \nonumber \\
\label{eqp35n}
\eear
and also
\bear
\tilde{\bf a}_i&=&(1+s \sigma_i/2) e^{-s \sigma_{i}}/s^2\nonumber \\
\tilde{\bf c}_i&=&e^{-s \sigma_{i}}/s\nonumber \\
\eear
The pair correlation function  is
\bear
2 \pi {\tilde g}_{ij}(s)=-\frac{e^{-s \sigma_{ij}}}{s ^2 D_\tau}\left\{1+s\left[(Q^{HS})'_{ij}(\sigma_{ij})\right]\right\}\,\,\,\,\,\,\,\,& &\nonumber \\
D_\tau
=1-\sum_j\rho_j \sigma_j^2\left \{\sigma_j\psi_1(s \sigma_j)A^0_j-\phi_1(s \sigma_j)\beta^0_j +
\frac{4\pi^2}{\Delta^2}\sum_{i}\rho_i \sigma_i^3 \psi_1(s \sigma_i)  \phi_1(s \sigma_j)\lambda_{ji}\right\}\,\,\,\,& &\nonumber \\
\eear

\subsection{The electrolyte limit}
We take the limit \cite{b80}
\be
z_m=0
\ee
of the general definitions. Then, 
\bear
a_i&=&\rho_i \sigma_i^3 \psi_1(s \sigma_i)\qquad b_j=A_j^0+\frac{\pi }{\Delta}a^0_j P_n =A_j\nonumber \\
c_i&=&\rho_i \sigma_i^2 \phi_1(s \sigma_i)\qquad d_j=\beta_j^0+ a_j^0 \Delta_N =\beta_j\nonumber \\
e^{(0)}_i&=&\rho_i e^{s\sigma_i/2}\sigma_i^2 \varphi^{(0)}_1(s \sigma_i)\qquad f^{(0)}_j=a^{(0)}_je^{-s \sigma_{j}/2}
\nonumber \\
 \varphi^{(0)}_1(s \sigma_i)&=& \frac{1}{s}\left\{{\pi\over\Delta}P_n \phi_1(s\sigma_i)-X_i^{(0)}[1+\phi_0(s\sigma_i)]\right\}\nonumber \\
\label{eq41n}
\eear
Moreover
\bear 
\tilde{q}^{0^{'}}_{ij}(i s)& =&[(1+s \sigma_i/2) A_j+s \beta_j]e^{-s \sigma_{ij}}/s^2-
\sum_m \frac{z_m}{s+z_m} e^{-(s+z_m) \sigma_{ij}}C^{(m)}_{ij}\nonumber \\
& =&{e^{-s \sigma_{ij}}\over s^2}\left\{[(1+s \sigma_i/2) A_j+s \beta_j]+
s{\mathcal B}_i^{(0)}a_j^{(0)}\right\}\nonumber \\
& =&{e^{-s \sigma_{ij}}\over s^2}\left\{[(1+s \sigma_i/2) A^0_j+s \beta^0_j]+
s{\mathcal X}_i^{(0)}a_j^{(0)}\right\}\nonumber \\
\eear
and also
\bear
\tilde{\bf a}_i&=&{(1+s \sigma_i/2)\over s^2} e^{-{s \sigma_{i}\over2}}\nonumber \\
\tilde{\bf c}_i&=&{e^{-{s \sigma_{i}\over2}}\over s}\nonumber \\
\tilde{\bf e}^{(0)}_i&=&\frac{z_m}{s+z_m} e^{-(s) \sigma_{i}/2}e^{-z_m \sigma_{i}}(\hat{\cal B}_i^{(m)}-\delta_i^{(m)})\Rightarrow\frac{1}{s} e^{-s \sigma_{i}/2}{\cal B}_i^{(0)}\nonumber \\
\label{eq:p20}
\eear
so that from eq.(\ref{eq35n}) we get the symmetric expression (compare \cite{bh78})
\bear
2 \pi {\tilde g}_{ij}(s)&=&-\frac{\widetilde \mu_{ij}}{D_\tau}
\nonumber \\
&=&-{e^{-s \sigma_{ij}}\over s^2 D_\pm}\left\{1+s \left[{\sigma_i\over2 } A^0_j+ \beta^0_j+
{\mathcal X}_i^{(0)}a_j^{(0)}\right]\right\}
\nonumber \\
\label{eq44n}
\eear
with\\
\bear
D_\pm=Det
\left| \begin{array}{ccc} 1-(ab)&-(ad)&-(af)\\ -(cb)&1-(cd)&-(cf)\\-(eb) &-(ed)&1-(ef) \end{array} \right|
\eear
where the scalar products ( see also \cite {bh78}) are defined by
\bear
(ab)=\sum_i a_ib_i,\quad(cd)=\sum_i c_id_i,\quad (ef)=\sum_i e_if_i\quad ....
\eear
and we use the definitions of eq.(\ref{eq41n}).

\section{ THE PRIMARY CLOSURE}
\vspace{0.5cm}

 The MSA closure condition obtained from Eq.(\ref{eq:msac}) is
\begin{equation}
2\pi K\delta_{i}^{(n)}\delta_{j}^{(n)}/z_{n}=\sum_{\ell}D_{i{\ell}}%
^{(n)}[\delta_{{\ell} j}-\rho_{\ell}\tilde{q}_{j{\ell}}(iz_{n})]\label{eq:p17}%
\end{equation}
Using the results of the last section we got  \cite{bluhe02}
\newpage
\bear
& & 2\pi K\delta_{j}^{(n)}
+z_{n}\sum_{\ell}
\mathcal{I}_{j\ell }^{(n)}a_{\ell}^{(n)}\nonumber \\
&-&\sum_{m}\frac{z_n}{z_{n}+z_{m}}\{\sum_{k}\rho_{k}a_{k}^{(n)}a_{k}^{(m)}\}
\left[  \sum_{\ell}\mathcal{J}_{j\ell}^{(n)}[\Pi_{\ell}^{(m)}-z_{m}X_{\ell
}^{(m)}]-\mathcal{I}_{j\ell}^{(n)}X_{\ell}^{(m)}\right]  =0\label{eqp46}
\nonumber \\
\eear

\section{The 1 Component Case: An explicit continued fraction solution}\label{1comp} 

In the one component case eq.(\ref{eqp46}) is simply \cite{bluhe02}
\be
-2\pi K_{n}\rho \left[ X_n\right] ^{2}
%=\frac{\partial (\Delta S/k)}{\partial \Gamma_n}
=z_{n}\beta _n\left[ 1+\sum_{m}%
\frac{1}{z_{n}+z_{m}}\beta _m\right] 
\label{eq91a}
\end{equation}
with
\be
X_n=\frac{\delta _n}{{\cal I}^{(n)}+\Gamma_n {\cal J}^{(n)}}
\label{1cc12} 
\end{equation}
\be
 \beta _n=\rho a_n X_n
\ee
In the diagonal approximation  we get explicit equations for the scaling parameters $\beta_i$ \cite{blmub}
( See appendix III.,\cite{blmub})
We recall that
\be 
{\cal J}^{(n)}=\sigma\phi_0(z_n\sigma)
+\rho\Delta^{J,(n)}
\label{eq:s12fa}
\end{equation}
\be 
\Delta^{J,(n)}=
-\frac{2\pi}{\Delta}  \sigma^4\psi_1( z_n\sigma)
\label{eqs12fab}
\end{equation}
and
\be 
{\cal I}^{(n)}=1+\rho  \Delta^{I,(n)}
\label{eq:s13fa}
\end{equation}
\be
\Delta^{I,(n)}=\sigma^2\phi_0(z_n\sigma) \frac{\pi}
{2\Delta}  \sigma-
\frac{2\pi}{\Delta}  \sigma^3\psi_1( z_n\sigma)
\left[ 1+\left(\zeta_2 \frac{\pi}
{2\Delta}\right)\sigma +\sigma z_n/2 \right]
\label{eqs13fa}
\end{equation}
and  we have used the definition (\ref{eq:s13fb}).\\

The solution of the $\beta$ equations is obtained solving the linear equation \cite{blmub}
\bear
 \overrightarrow{\beta }=\left[\hat{\cal M}\right]^{-1}\cdot 2  \overrightarrow{\Gamma } \nonumber \\
\label{bteq}
\eear
where\\
\be
\hat{\cal M}= \left[ 
\begin{array}{cccc}
1 & 1-\gamma _{12} &1-\gamma _{13}  &.\\ 
1+\gamma _{12}  & 1  &1-\gamma _{23}  &.\\
1+\gamma _{13} &1+\gamma _{23} &1 &.\\
.&.&.&.
\end{array}
\right]
 ;\qquad
{\overrightarrow{\beta }}= \left[ 
\begin{array}{c}
\beta_1 \\ 
\beta_2  \\
.\\ 
\end{array}
\right];\qquad
{\overrightarrow{\Gamma}}=\left[ 
\begin{array}{c}
\Gamma_1 \\ 
\Gamma_2 \\
. 
\end{array}
\right]
\label{mcal}
\end{equation}

and

\be
\gamma _{nm}=\frac{2\Gamma _{n}+z_{n}-2\Gamma _{m}-z_{m}}{z_{m}+z_{n}} 
\label{gdef}
\ee

We give the soultions for the first few cases \cite{blmub}:\\
\begin{enumerate}
\item 1 Yukawa $\beta_1=2 \Gamma_1$

\item 2 Yukawas

\be
\beta_1=\frac{2\Gamma_2-\beta_s}{\gamma_{12}}\qquad 
\beta_2=\frac{-2\Gamma_1+\beta_s}{\gamma_{12}}\qquad \beta_s= \frac{2\Gamma_2-2\Gamma_1}{\gamma_{12}}
\ee

\item 3 Yukawas:
This case is slightly more complicated:\\
The  resolvent is
\be
\overrightarrow{{\mu}}^{(3;1)}
=\left(\begin{array}{c}
\gamma_{23}\\
\gamma_{31}\\
\gamma_{12}
\end{array}\right),
\ee
so that the explicit solution is
\[
\beta _{1}=\left[\frac{1}{d_3}\right]\left[\gamma _{23}\left(\frac{2\overrightarrow{\Gamma }\cdot {\overrightarrow{{\mu}}^{(3;1)}}}{d_3}+s_{23}\right)-z_{23}\right]
\]
\[
\beta _{2}=-\left[\frac{1}{d_3}\right]\left[\gamma _{13}\left(\frac{2\overrightarrow{\Gamma }\cdot {\overrightarrow{{\mu}}^{(3;1)}}}{d_3}+s_{13}\right)-z_{13}\right]
\]
\be
\beta _{3}=\left[\frac{1}{d_3}\right]\left[\gamma _{12}\left(\frac{2\overrightarrow{\Gamma }\cdot {\overrightarrow{{\mu}}^{(3;1)}}}{d_3}+s_{12}\right)-z_{12}\right]\label{eq5}
\end{equation}
where $d_3=\gamma_{12}+\gamma_{23}-\gamma_{13}$.
\be
\beta_s=\frac{2\overrightarrow{\Gamma }\cdot {\overrightarrow{{\mu}}^{(3;1)}}}{d_3}
\label{bsss}
\ee

\end{enumerate}
The general case of $n\ge3$ can be found in the work of Blum et al. \cite {blmub,blmub1}

\subsection{The 1 Yukawa case}
In the 1-Yukawa, 1 component case  { [27]} the closure equation  is simply
\be
-2 \pi \rho K_1 [X_1]^2 =z_1 \beta_1 \left[1+ \frac{\beta_1}{2 z_1}\right]
;\qquad
\ee
so that putting it all together we get the equation
\be
y_1=2\Gamma_1(\Gamma_1+z_1)\left[{\cal I}^{(1)}+\Gamma_1 {\cal J}^{(1)}\right]^2
;\qquad
y_1=-\frac{2\pi \rho K_1\delta _1^2}{z_1}
\ee
\bigskip 
The  physical branch  yields the recursion relation
\be
\Gamma_1^{(1)}=\frac{y_1}{ 2 z_1\left[{\cal I}^{(1)}\right]^2}
;\qquad
\Gamma_1^{(n+1)}=\frac{y_1}{2(\Gamma_1^{(n)}+z_1)\left[{\cal I}^{(1)}+\Gamma_1^{(n)} {\cal J}^{(1)}\right]^2}
\label{eqy4}
\ee

For large $z_1 \geq 3$, we may disregard the exponential terms in Eqs(\ref{eq:s13fb}), and then we get the simple

\be
{\cal I}^{(1)}=1 + \frac{2\,\pi \,\rho \,
     \left( 1 + \frac{\pi \,\rho }{3} - \frac{\pi \,z_1\,\rho }{4} \right) }{z_1^3\,{\Delta }^2};\qquad
{\cal J}^{(1)}=\frac{1}{z_1} + \frac{\pi \rho \left( -2 + z_1 \right) \, }{z_1^3\,\Delta }
\ee
from where
\be
 \Gamma_1=\frac{y_1}
  {2 z_1{\left( 1 - \frac{{\pi }^2\,{\rho }^2}{2\,z_1^2\,{\Delta }^2} + 
        \frac{2\,\pi \,\rho \,\left( 3 + \pi \,\rho  \right) }{3\,z_1^3\,{\Delta }^2} \right) }^2};\qquad
\Gamma_2=\frac{y_1}{2(\Gamma_1+z_1){\cal D}_{\Gamma,1}^2}
\ee
with
\[
{\cal D}_{\Gamma,1}= 
1+ \frac{1}{z_1^2\,\Delta }\left[{\frac{-\left( {\pi }^2\,{\rho }^2 \right) }{2\Delta} + 
        \frac{2\,\pi \,\rho \,\left( 1 + \frac{\pi \,\rho }{3} \right) }{z_1 \Delta} - 
     \frac{2\,y_1\,\left( - \Delta+
          \frac{2\pi \rho \,\left( 2 -   z_1 \right)  }{z_1^2} \right) }
        {\,{\left( 2 - \frac{{\pi }^2\,{\rho }^2}{z_1^2\,{\Delta }^2} + 
            \frac{4\,\pi \,\rho \,\left( 1 + \frac{\pi \,\rho }{3} \right) }{z_1^3\,{\Delta }^2}
            \right) }^2}}\right]
\]

 The solution is always convergent  for the atractive case, and we do get  the Onsagerian limits for large densities and zero temperature correctly. It has been successfully  tested numerically against other numerical methods \cite{vazquez03} and Hernando, (unpublished, but using Pastore's criterion). For the  repulsive case $K>0$ this equation has poles and  unphysical divergencies occurring when the Onsagerian limits are attained. These divergencies are removed when a suitable reference hard core is introduced. \\
\subsection{ The multiyukawa case}
The above solution is valid for any number of Yukawas $n\ge 1$. We write Eq. (\ref{eq91a}) in matrix form:

\be
 \left[ 
\begin{array}{c}
y_1 \\ 
y_2  \\
.\\ 
\end{array}
\right]=
\left[ 
\begin{array}{ccc}
1+\beta_1/s_{11} & \beta_1/s_{12}&.\\ 
\beta_2/s_{12} & 1+\beta_2/s_{22}  &.\\
.&.&.
\end{array}
\right]\left[ 
\begin{array}{c}
\beta_1 \\ 
\beta_2 \\
. 
\end{array}
\right]
\end{equation}
Using equations (\ref{bteq}) and (\ref{mcal})
we find
\be
\overrightarrow{\Gamma}=\frac{1}{2}\hat{\cal M}\cdot\left[ 
\begin{array}{ccc}
1+\beta_1/s_{11} & \beta_1/s_{12}&.\\ 
\beta_2/s_{12} & 1+\beta_2/s_{22}  &.\\
.&.&.
\end{array}
\right]^{-1}\cdot \left[ 
\begin{array}{c}
y_1/[{\cal I}^{(1)}+\Gamma_1 {\cal J}^{(1)}]^2
 \\ 
y_2/[{\cal I}^{(2)}+\Gamma_2 {\cal J}^{(2)}]^2  \\
.\\ 
\end{array}
\right]
\label{myuki1}
\ee 
with
\[
s_{nm}=z_n+z_m
\]
This relation is the extension of Eq.(\ref{eqy4}) to the multiyukawa case: The iteration follows exactly 
the same steps, only that now we have to use
the extra equation (\ref{bteq}):
The first iterate is
\bear
\overrightarrow{\Gamma}^{(0)}=\frac{1}{2}
 \left[ 
\begin{array}{cccc}
1 & 1-\frac{z_{12}}{s_{12}} &1-\frac{z_{13}}{s_{13}} &.\\ 
1+\frac{z_{12}}{s_{12}}  & 1  &1-\frac{z_{23}}{s_{23}}  &.\\
1+\frac{z_{13}}{s_{13}}  &1+\frac{z_{23}}{s_{23}} &1 &.\\
.&.&.&.
\end{array}
\right]
\cdot \left[ 
\begin{array}{c}
y_1/[ {\cal I}^{(1)}]^2
 \\ 
y_2/[ {\cal I}^{(2)}]^2  \\
.\\ 
\end{array}
\right];\quad z_{nm}=z_n-z_m \nonumber \\
\label{myuki2}
\eear
and then
\be
\overrightarrow{\Gamma}^{(n+1)}=\frac{1}{2}\hat{\cal M}^{(n)}
\cdot\left[\left\{ 
\begin{array}{ccc}
1+\beta_1/s_{11} & \beta_1/s_{12}&.\\ 
\beta_2/s_{12} & 1+\beta_2/s_{22}  &.\\
.&.&.
\end{array}\right\}^{(n)}
\right]^{-1}\cdot \left[ 
\begin{array}{c}
y_1/[{\cal I}^{(1)}+\Gamma_1^{(n)} {\cal J}^{(1)}]^2
 \\ 
y_2/[{\cal I}^{(2)}+\Gamma_2 ^{(n)}{\cal J}^{(2)}]^2  \\
.\\ 
\end{array}
\right]
\label{myuki}
\ee 
where the superscript $(n)$ indicates the level of the 
iteration.\\
\section {Summary of results}
\begin {itemize}
\item A new formulation of the pair distribution function for a large class of systems represented by a multiyukawa closure of the Ornstein-Zernike equation (\ref{eq1}) which includes systems such as water, ionic mixtures, ion channels and polymers is presented. This formulation is considerably simpler and more explicit than those in the literature because it involves directly measurable parameters such as the contact pair distribution function. It is also capable of handling the new improved versions of the MSA and the realistic octupolar model of water \cite{bluvede}.
\item The new result shown in equation (\ref{eq2}) is simple and explicit:
For example in the case of the SANS and SAXS experiments \cite{chen05,broccio06} our result show explicitly the connection between the number of peaks in the diffraction spectra to the theoretical formula. 
because we get for dilute systems a sum of M Lorentzians, one for each yukawa term in the closure.
 \item The one component case has been discussed in several papers, largely in collaboration with J. Hernando \cite{bhe01}, although most of the numerical results are still unpublished. The conclusion is that once the branch points are identified the continued fraction formalism {\it always} converges ( see also\cite{vazquez03}).
\end {itemize}
\section{Appendix I: The algebraic solution of the general Yukawa closure of the Ornstein Zernike equation}
We quote the results from the review of  Blum and Hernando \cite{bluhe02}: 
The solution of the system of equations  (\ref{eq:p1},\ref{eq22d}) yields 
\begin{equation}
A_{j}=A_{j}^{0}+\frac{\pi}{\Delta}\sum_{n}a_{j}^{(n)}P^{(n)}\label{eq:s6b};\qquad
\beta_{j}=\beta_{j}^{0}+\frac{2\pi}{\Delta}\sum_{n}a_{j}^{(n)}\Delta
^{(n)}\label{eq:p3b}%
\end{equation}
where
\begin{equation}
A_{j}^{0}=\frac{2\pi}{\Delta}\left[  1+(1/2)\zeta_{2}\frac{\pi}{\Delta}%
\sigma_{j}\right] \label{eqs7ab};\qquad
\beta_{j}^{0}=\frac{\pi}{\Delta}\sigma_{j}%
\end{equation}
\begin{equation}
\frac{\pi}{\Delta}P^{(n)}=\frac{1}{z_{n}}\sum_{\ell}\rho_{\ell}\left[
A_{\ell}^{0}X_{\ell}^{(n)}+2\beta_{\ell}^{0}(z_{n}X_{\ell}^{(n)}-\Pi_{\ell
}^{(n)})\right] \label{A1}%
\end{equation}
\begin{equation}
\Delta^{(n)}=-\frac{1}{z_{n}^{2}}\sum_{\ell}\rho_{\ell}\left[ X_{\ell}^{(n)}
(A_{\ell}^{0}-z_{n}\beta_{\ell}^{0})+2\beta_{\ell}^{0}(z_{n}X_{\ell}^{(n)}%
-\Pi_{\ell}^{(n)})\right] \label{A2}%
\end{equation}
We have
\bear
\Pi_{j}^{(n)}={\hat{B}}_{j}^{(n)}+\Delta^{(n)}+\frac{\pi}{2\Delta}\sigma
_{j}P^{(n)}\label{eq:s9c};\nonumber \\
X_{i}^{(n)}=\delta_{i}^{(n)}+\sigma_{i}{\hat{B}}_{i}^{(n)}\phi_{0}(z_{n}%
\sigma_{i})+\sigma_{i}\Delta^{(n)}\label{eq:fi7}
\eear
\bear
\Pi_{j}^{(n)}&=&\hat{\xi}_{j}^{(n)}+\sum_\ell \hat{\cal I}_{j\ell}^{(n)}{\hat{B}}_{\ell}^{(n)}\nonumber \\
X_{j}^{(n)}&=&\gamma_{j}^{(n)}
+\sum_\ell \hat{\cal J}_{j\ell}^{(n)}{\hat{B}}_{\ell}^{(n)}
\label{eqfi7} 
\eear
where
 \bear
\hat{\cal I}_{j\ell }^{(n)}
&=&\delta _{j\ell }-\rho _{\ell}{\sigma_{\ell}^2}
\lbrack \beta^0_j
\phi_1(z_n \sigma_\ell)+A^0_j
\sigma_{\ell}\psi_1(z_n \sigma_\ell)\rbrack\nonumber \\
\hat{\cal J}_{j\ell }^{(n)}&=&\delta _{j\ell }
\sigma_j\phi _{0}(z_{n}\sigma_{\ell})
-2\rho _{\ell }\beta^0_j\sigma_{\ell}^3\psi _{1
}(z_{n}\sigma_{\ell})
\label{eq:s13fb} 
\eear
with
\bear
\hat{\xi}_{j}^{(n)}=-\frac{1}{z_n^2}\sum_\ell\rho_\ell\delta_{\ell}^{(n)}\lbrack A^0_j +z_n Q'(\sigma_{\ell j})\rbrack\nonumber \\
\hat{\gamma}^{(n)}_j=\delta^{(n)}_j-\frac{2 \pi \sigma_j}{z_n^2\Delta}\sum_\ell \rho_\ell \delta^{(n)}_\ell\left(1+\frac{z_n\sigma
_{\ell}}{2}\right)
\label{xigama} 
\eear
Here we define \cite{bluhe02}:
\be
\zeta _{n}=\sum_{k}\rho _{k}\sigma _{k}^{n}\label{eq:p6} 
\ee
\be
\Delta =1-\pi \zeta _{3}/6\label{eq:p7} 
\ee
\be
{\tilde{g}}_{ij}(s)=\int_{0}^{\infty }drrg_{ij}(r)e^{-sr}\label{eq:p11} 
\ee
\be
\gamma _{ij}^{(n)}=2\pi {\tilde{g}}_{ij}(z_{n})\rho _{j}/z_{n}\label{eq:p8} 
\ee
\be
{\hat{B}}_{j}^{(n)}=\sum_{i}z_{n}\delta _{i}^{(n)}\gamma
_{ji}^{(n)}e^{z_{n}\sigma _{ij}}\label{eq:fi1} 
\ee
\bear
\psi_1( s\sigma_i)=\frac{1}{2}[\gamma_z(3,s\sigma_i)-\gamma_z(2,s\sigma_i)]
\nonumber \\
\phi_1( s\sigma_i)=[\gamma_z(2,s\sigma_i)-\gamma_z(1,s\sigma_i)]
\nonumber \\
\eear
with the incomplete $\gamma$ function
\bear
\gamma_z(n,z)=\frac{(n-1)!}{z^n}\left[1-e^{-z}\sum_i^{n-1}\frac{z^i}{i!}\right]
\eear
\section {Appendix II: Inverse of ${\mathbf M}_{ij}(s)$}
\bear
{\widehat{\mathbf M }_{ij}{(s)}}
&=&\delta_{ij}-\mu_{ij}
;\qquad
\mu_{ij}=\sum_{\alpha=1}^{\alpha_m}{ a_i^{(\alpha)}}{ b_j^{(\alpha)}}
\nonumber \\
\eear
The determinant of this matrix can be expanded in its minors $\mathcal M_{ij}$
\bear
D_T=Det|\delta_{ij}-\mu_{ij}|=\sum_i [\delta_{ij}-\mu_{ij}]\,(-)^{i+j}{\mathcal M_{ij}}\quad;\qquad \forall j
\nonumber \\
\label{eq29}
\eear 
This determinant can also be condensed into a simpler and more compact form
\bear
D_T=D_\tau=\left|\delta_{\alpha,\beta}-\left(\sum_i a_i^{(\alpha)} b_i^{(\beta)}\right)\right|
\eear
which can also be expanded in its minors ${\mathbf M}^{(\alpha,\beta)}$
\bear
D_\tau&=&\sum_\beta \left(\delta_{\alpha,\beta}-a^{(\alpha)}\bullet b^{(\beta)}\right) [-]^{\alpha+\beta} {\mathbf M}^{(\alpha,\beta)};\quad \forall \,\alpha
\nonumber \\
&=&\sum_\beta \left(\delta_{\alpha,\beta}-a^{(\alpha)}\bullet b^{(\beta)}\right)  {\mathcal A}^{(\alpha,\beta)};\quad {\mathcal A}^{(\alpha,\beta)}=[-]^{\alpha+\beta} {\mathbf M}^{(\alpha,\beta)}
\nonumber \\
\label{eq30}
\eear
Using  Cramer's rule then we find the general inverse
\bear
[{\mathbf M }^{-1}(s)]_{ij}&=&\delta_{ij}+\frac{1}{D_{\tau}}\sum_{ \beta}a_i^{(\alpha)} b_j^{(\beta)}\left\{{\mathcal{A}}^{(\alpha,\beta)}\right\};\quad \forall\{\alpha,\beta\}\nonumber \\
\eear
%We verify that
\bear
\sum_k\{\delta_{ik}-\mu_{ik}\}.[{\mathbf M }^{-1}(s)]_{kj}&=&\delta_{ij}\nonumber \\
\sum_k\{\delta_{ik}-\mu_{ik}\}.[\delta_{ij}+\frac{1}{D_{\tau}}\sum_{ \beta}({-})^{\alpha+\beta}a_k^{(\alpha)} b_j^{(\beta)}\left\{{\mathcal{M}}^{(\alpha,\beta)}\right\}]&=&\delta_{ij}\nonumber \\
\sum_k\left[\sum_\gamma a_i^{(\gamma)} b_k^{(\gamma)}\right].\left[\frac{1}{D_{\tau}}\sum_{ \beta}a_k^{(\alpha)} b_j^{(\beta)}\left\{{\mathcal{A}}^{(\alpha,\beta)}\right\}\right]&=&\mu_{ij}\nonumber \\
\eear
which means that we must satisfy
\bear
\mu_{ij}=\sum_{\gamma,\beta}a_i^{(\gamma)} b_j^{(\gamma)}
\left[\frac{1}{D_{\tau}}\sum_k b_k^{(\gamma)} a_k^{(\alpha)}\right]{\mathcal{A}}^{(\alpha,\beta)}
\eear
\bear
& &\sum_k\left( \mathbf{\mu}_{ik}[{\mathbf M }^{-1}(s)]_{ij}^{\alpha,\beta}\right)
=\sum_\alpha a_i^{(\alpha)}\sum_k b_k^{(\beta)}\left(-\frac{1}{D_{\tau}}\sum_{\gamma} a_k^{(\gamma)}b_j^{(\delta)}\left\{{\mathcal{A}}_{\gamma,\delta}\right\}\right);\qquad \forall \delta\nonumber \\
&=&\frac{1}{D_{\tau}}\sum_{ \beta}({-})^{\alpha+\beta}a_i^{(\alpha)} b_j^{(\beta)}\left\{{\mathcal{M}}^{(\alpha,\beta)}\right\}]+\sum_\alpha a_i^{(\alpha)}\left(-\frac{1}{D_{\tau}}\sum_{\gamma}\left[\sum_k b_k^{(\beta)} a_k^{(\gamma)}\right]\left\{{\mathcal{A}}_{\gamma,\delta}\right\}\right)b_j^{(\delta)};\nonumber \\
&=&\mathbf{\mu}_{ij}\nonumber \\
 \eear
as it should
\section{Acknowledgements} 
 Support from NSF
through grant DMR02-03755 and DOE grant DE-FG02-03ER 15422 is  acknowledged.\\
 We  wish Prof. Sow Hsin Chen a very happy and healthy 70th birthday. The subject of this paper was inspired in conversationss with him.\\
\bibliographystyle{plain}
\bibliography{} 

\end{document}